\documentclass[a4paper, 11pt]{article}
\usepackage[final]{pdfpages}
\usepackage{fancyhdr}
\usepackage[utf8]{inputenc}
\usepackage[protrusion=true,expansion=true]{microtype}
\usepackage[pagebackref=false, pdfpagelabels]{hyperref}
\hypersetup{
        pdftitle={LOFAR CRKSP Contributions 35th ICRC 2017},
        pdfproducer={pdflatex},
        colorlinks=true,
        linkcolor=black,
        citecolor=black,
        filecolor=black,
				urlcolor=black}

\usepackage{xcolor}
\definecolor{shadecolor}{RGB}{255,255,255}
\newcommand{\mybox}[1]{\par\noindent\colorbox{shadecolor}
{\parbox{2cm}{\centering~#1~}}}
\usepackage{geometry}
\usepackage{tocloft}

\let\oldsection\section
\renewcommand{\section}{\cleardoublepage\oldsection}%

\makeatletter
\DeclareRobustCommand\authortoctext[1]{%
{\addvspace{-3pt}\nopagebreak\leftskip0em\relax
\rightskip \@tocrmarg\relax
\noindent\itshape#1\par\addvspace{-14pt}}}
\makeatother

\begin{document}
\thispagestyle{empty}
{
	\hfill~\includegraphics[height=2cm]{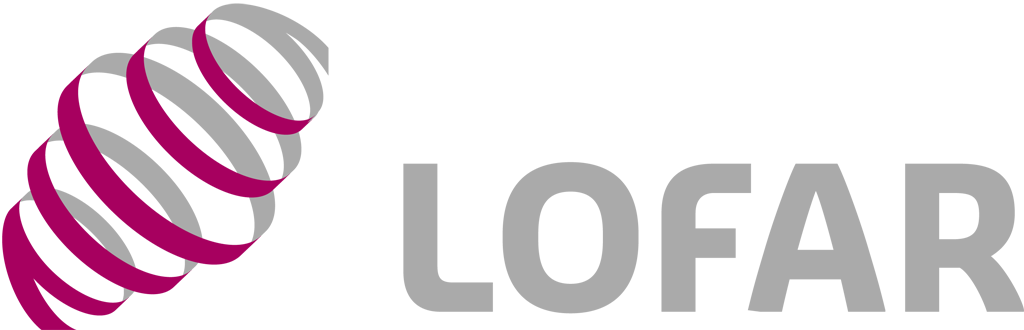} \\[16mm]
	\bfseries\huge\sffamily\noindent Contributions of the LOFAR Cosmic Ray Key Science Project to the 35th International Cosmic Ray Conference (ICRC 2017)\\[8mm]
}
{
	\noindent\bfseries\raggedright A.~Bonardi$^{1}$, S.~Buitink$^{2}$, A.~Corstanje$^{1}$,
	H.~Falcke$^{1,3,4}$, B.~M.~Hare$^{5}$, J.~R.~Hörandel$^{1,3}$,
	P.~Mitra$^{2}$, K.~Mulrey$^{2}$, A.~Nelles$^{1,6}$, J.~P.~Rachen$^{1}$,
	L.~Rossetto$^{1}$, P.~Schellart$^{1,7}$, O.~Scholten$^{5,8}$, S.~ter
	Veen$^{1,4}$, S.~Thoudam$^{1,9}$, T.~N.~G.~Trinh$^{5}$, T.~Winchen$^{2}$\\[4mm]
}
{
\noindent
$^1$ Department of Astrophysics/IMAPP, Radboud University Nijmegen, P.O. Box 9010, 6500 GL, Nijmegen, The Netherlands,\\
$^2$ Astrophysical Institute, Vrije Universiteit Brussel, Pleinlaan 2, 1050 Brussels, Belgium,\\
$^3$ NIKHEF, Science Park Amsterdam, 1098 XG Amsterdam, The Netherlands,\\
$^4$ Netherlands Institute of Radio Astronomy (ASTRON), Postbus 2, 7990 AA Dwingeloo, The Netherlands,\\
$^5$ KVI-CART, University Groningen, P.O. Box 72, 9700 AB Groningen,\\
$^6$ Department of Physics and Astronomy, University of California Irvine, Irvine, CA 92697-4575, USA,\\
$^7$ Department of Astrophysical Sciences, Princeton University, Princeton, NJ 08544, USA,\\
$^8$ Interuniversity Institute for High-Energy, Vrije Universiteit Brussel, Pleinlaan 2, 1050 Brussels, Belgium,\\
$^9$ Department of Physics and Electrical Engineering, Linn\'euniversitetet, 35195 V\"axj\"o, Sweden
}
\vfill

\newpage
\pagestyle{fancy}
\tableofcontents

\newpage
\cleardoublepage
\phantomsection
\addcontentsline{toc}{section}{Cosmic ray mass composition with LOFAR}\includepdf[pages=1,pagecommand=\thispagestyle{empty}]{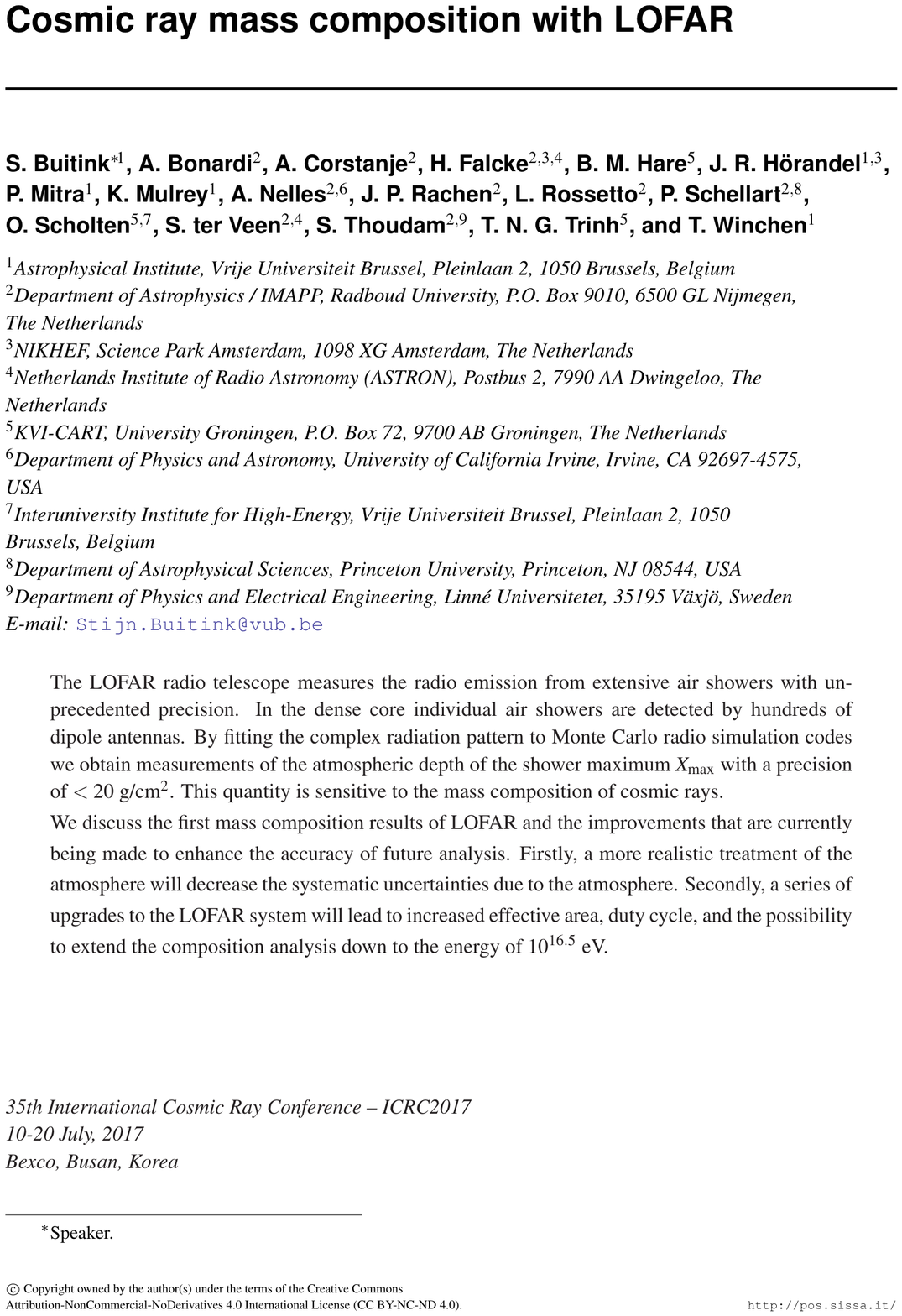}
\includepdf[pages=2-,pagecommand=\thispagestyle{fancy}]{ICRC2017_0499.pdf}

\cleardoublepage
\phantomsection
\addcontentsline{toc}{section}{Characterisation of the radio frequency spectrum
emitted by\newline high energy air showers with LOFAR }
\includepdf[pages=1,pagecommand=\thispagestyle{empty}]{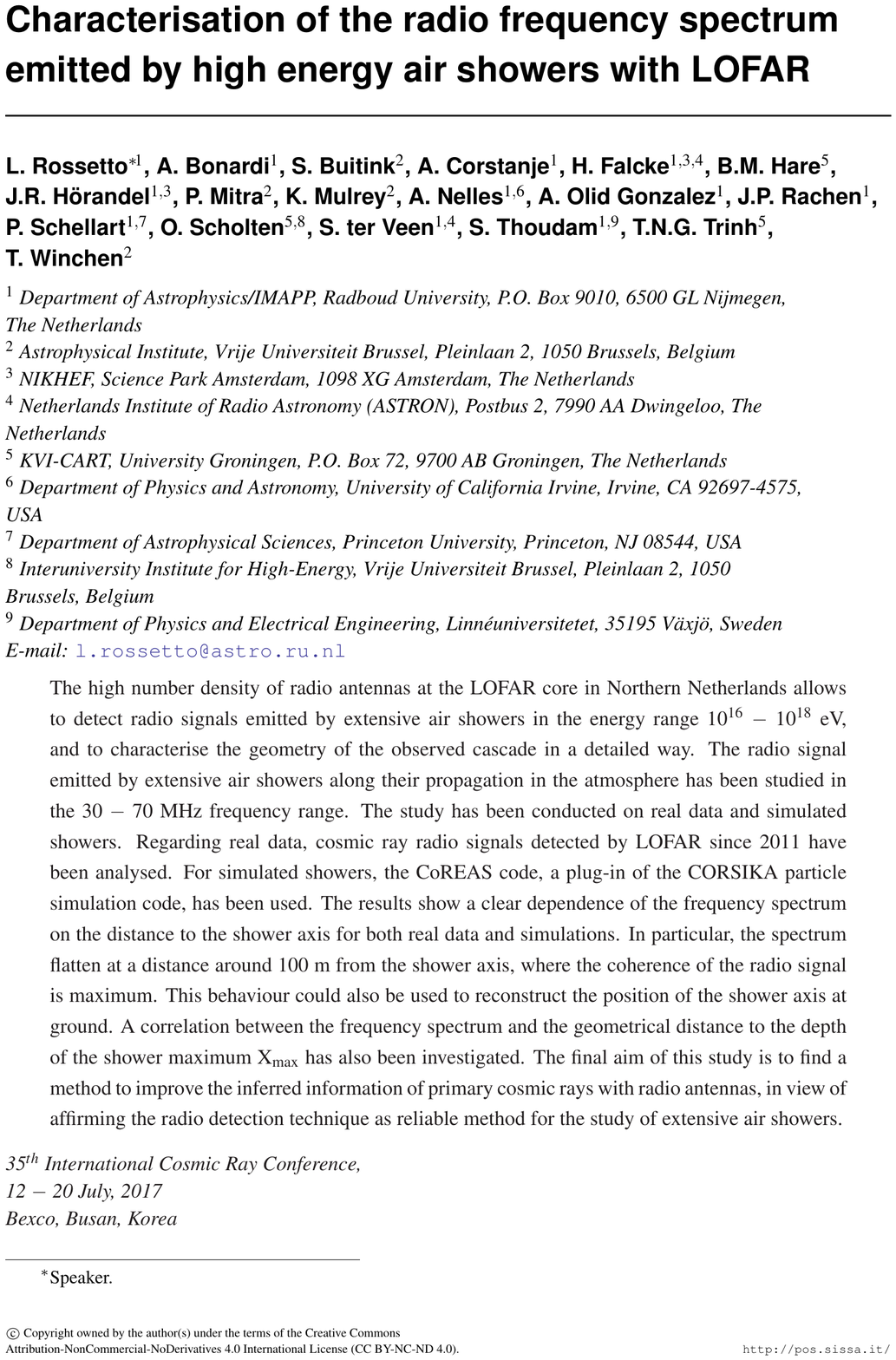}
\includepdf[pages=2-,pagecommand=\thispagestyle{fancy}]{ICRC2017_0329.pdf}

\cleardoublepage
\phantomsection
\addcontentsline{toc}{section}{The effect of the atmospheric refractive index on the radio signal of extensive air showers using Global Data Assimilation System (GDAS)}
\includepdf[pages=1,pagecommand=\thispagestyle{empty}]{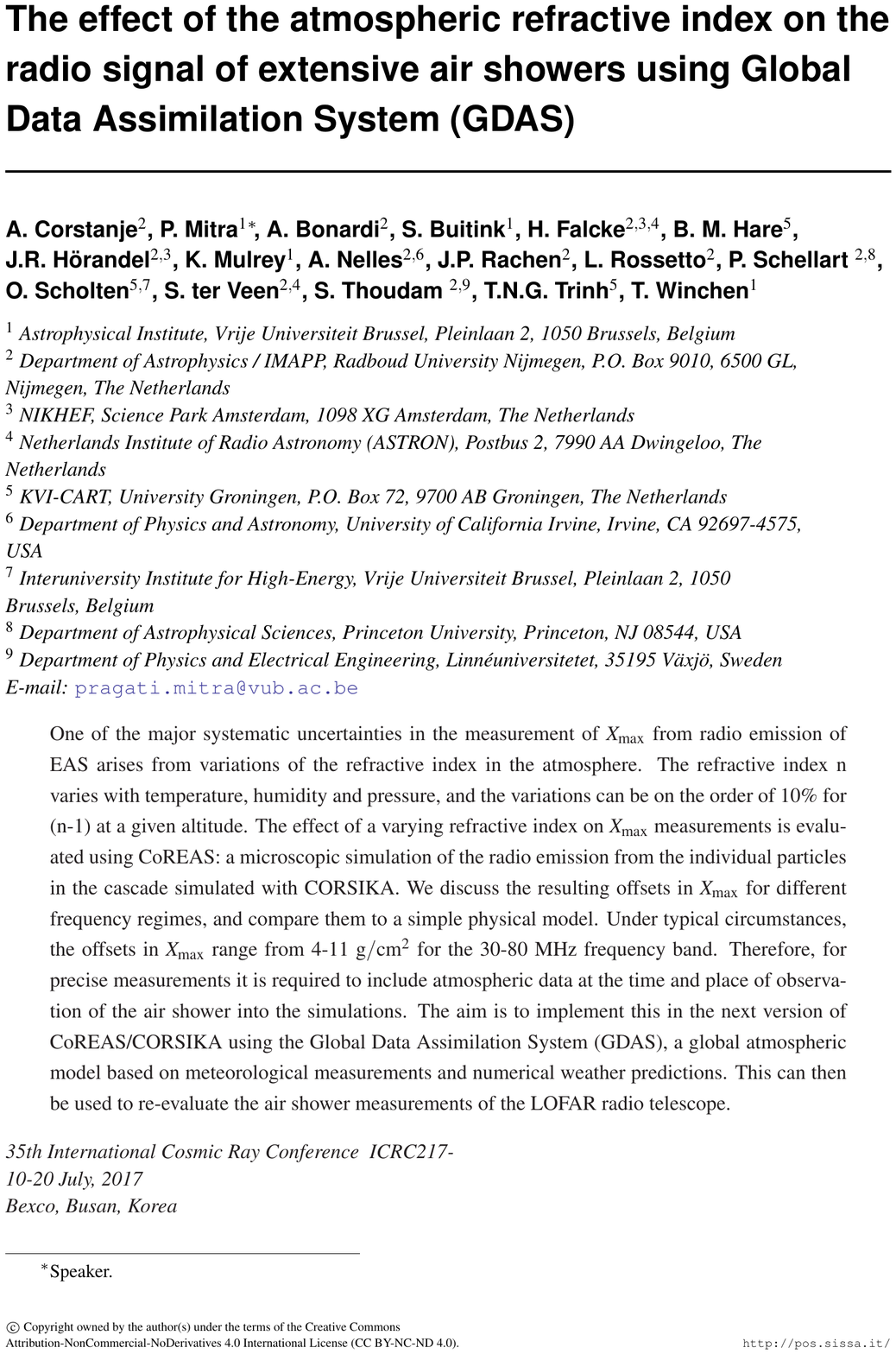}
\includepdf[pages=2-,pagecommand=\thispagestyle{fancy}]{ICRC2017_0325.pdf}

\newpage~\newpage
\cleardoublepage
\phantomsection
\addcontentsline{toc}{section}{Circular polarization in radio emission from extensive air showers}
\includepdf[pages=1,pagecommand=\thispagestyle{empty}]{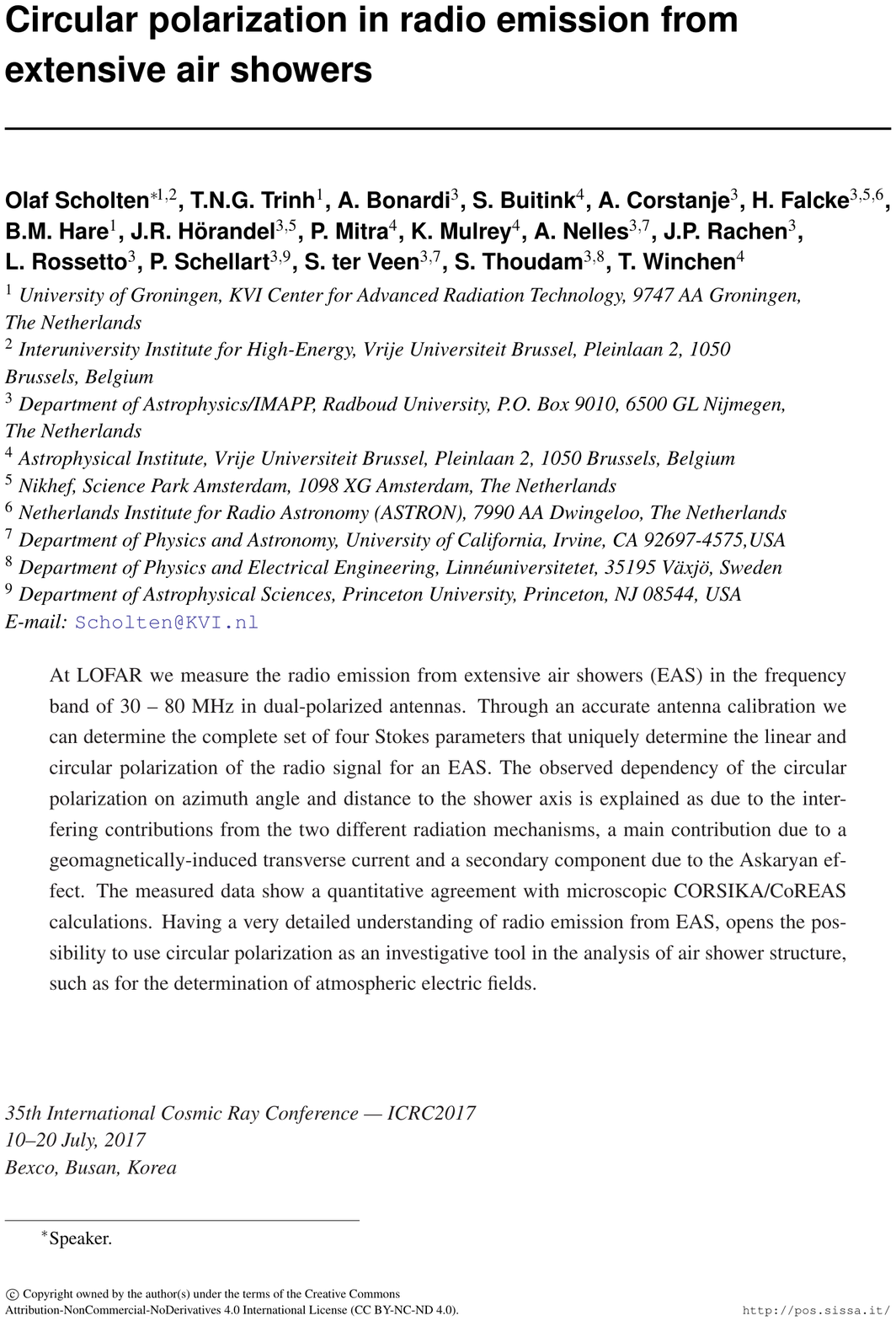}
\includepdf[pages=2-,pagecommand=\thispagestyle{fancy}]{ICRC2017_0324.pdf}

\cleardoublepage
\phantomsection
\addcontentsline{toc}{section}{Circular polarization of radio emission from
extensive air showers\newline probes atmospheric electric fields in thunderclouds.}
\includepdf[pages=1,pagecommand=\thispagestyle{empty}]{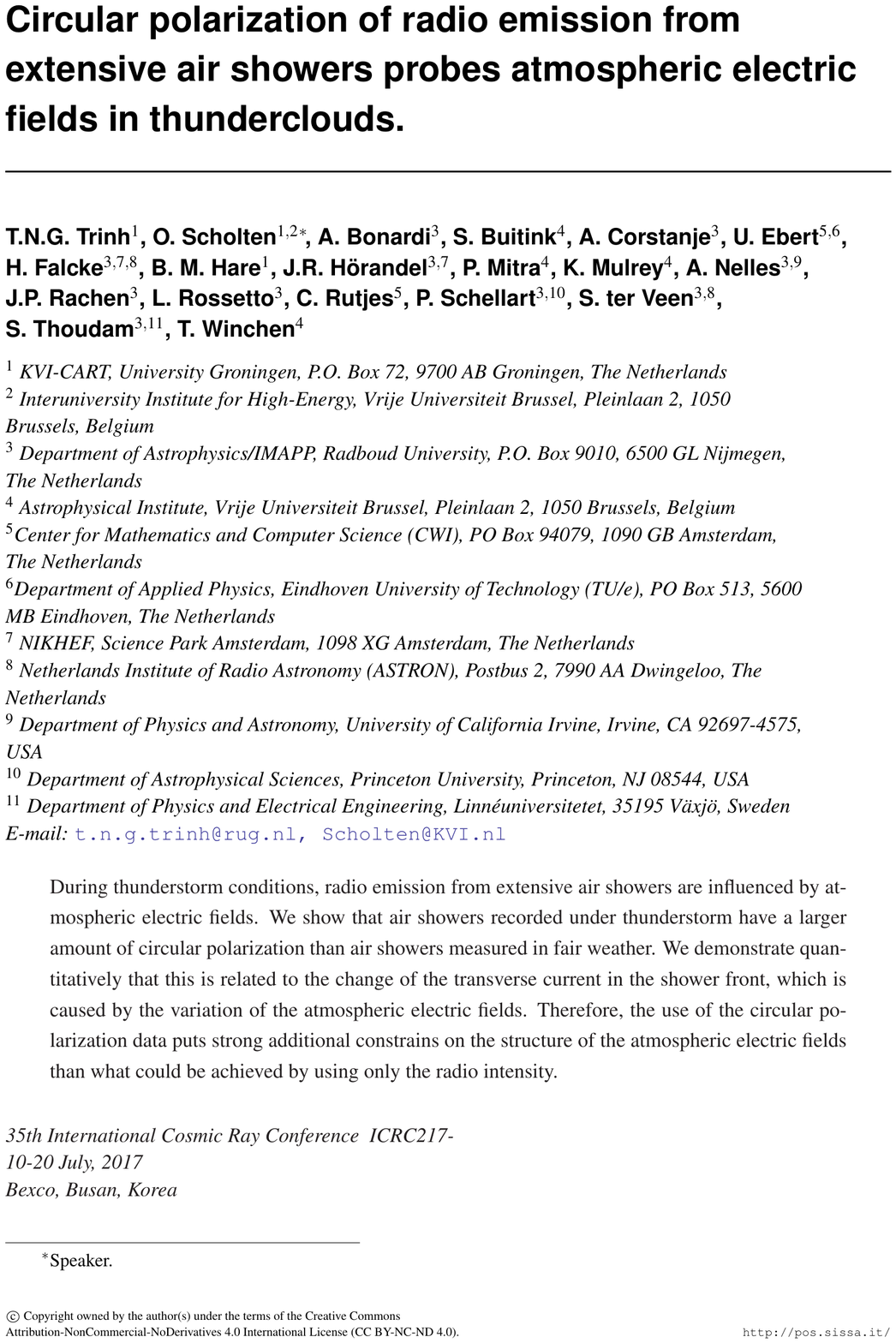}
\includepdf[pages=2-,pagecommand=\thispagestyle{fancy}]{ICRC_GiaTrinh.pdf}

\cleardoublepage
\phantomsection
\addcontentsline{toc}{section}{Study of the LOFAR radio self-trigger and single-station\newline acquisition mode}
\includepdf[pages=1,pagecommand=\thispagestyle{empty}]{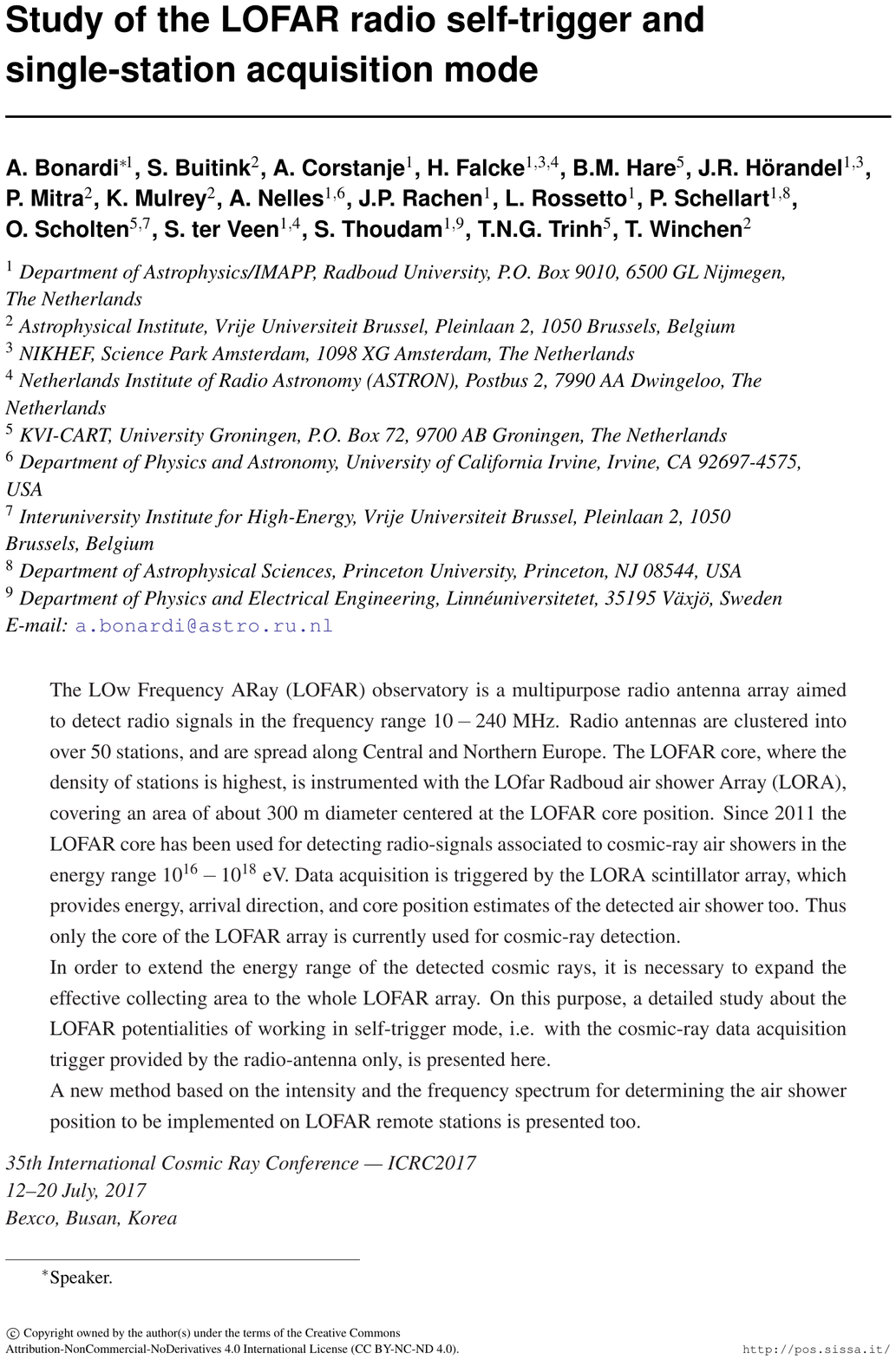}
\includepdf[pages=2-,pagecommand=\thispagestyle{fancy}]{ICRC2017_0402.pdf}

\cleardoublepage
\phantomsection
\addcontentsline{toc}{section}{Expansion of the LOFAR Radboud Air Shower Array}
\includepdf[pages=1,pagecommand=\thispagestyle{empty}]{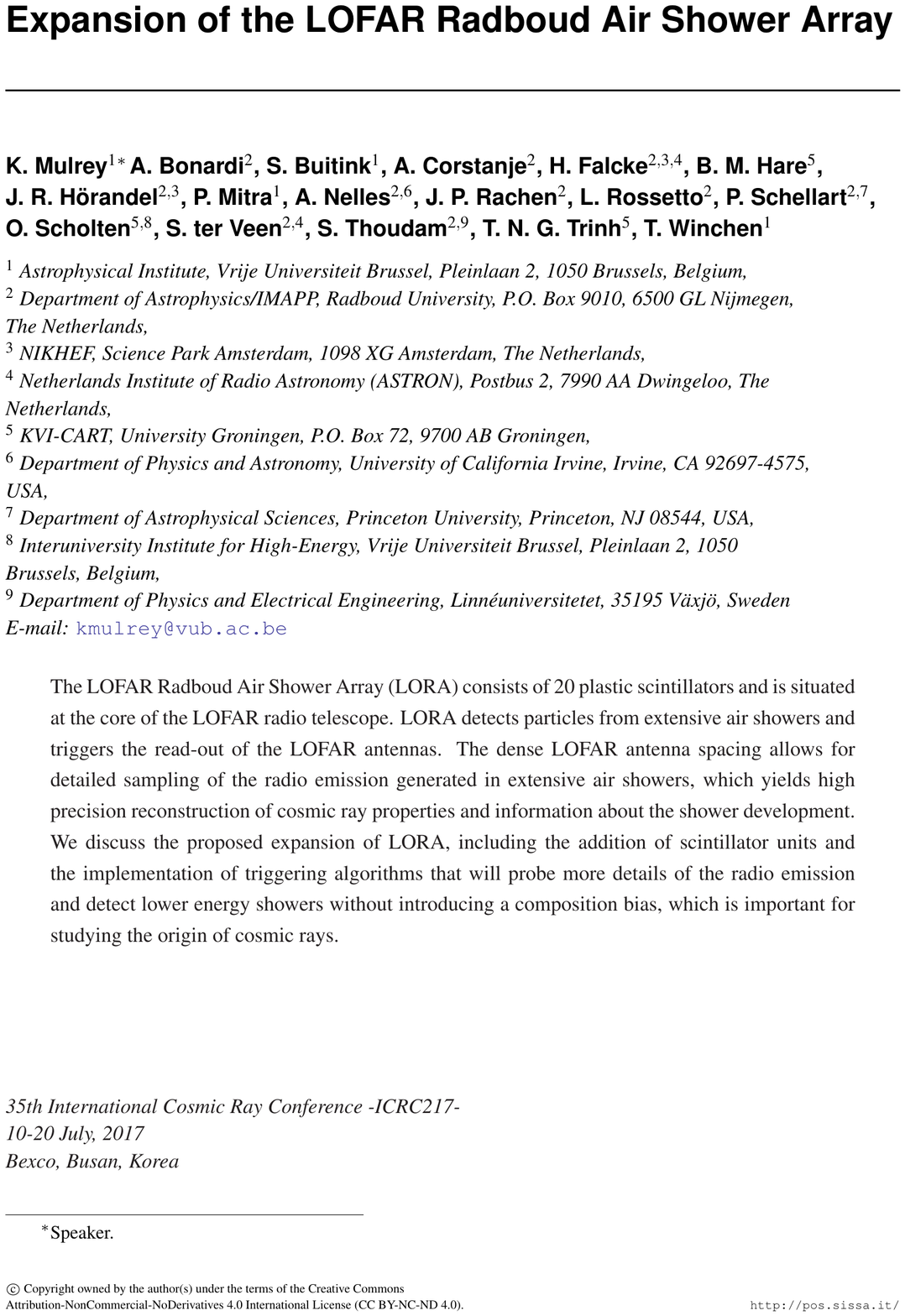}
\includepdf[pages=2-,pagecommand=\thispagestyle{fancy}]{ICRC2017_0413.pdf}

\cleardoublepage
\phantomsection
\addcontentsline{toc}{section}{Overview and Status of the Lunar Detection of\newline Cosmic Particles with LOFAR}
\includepdf[pages=1,pagecommand=\thispagestyle{empty}]{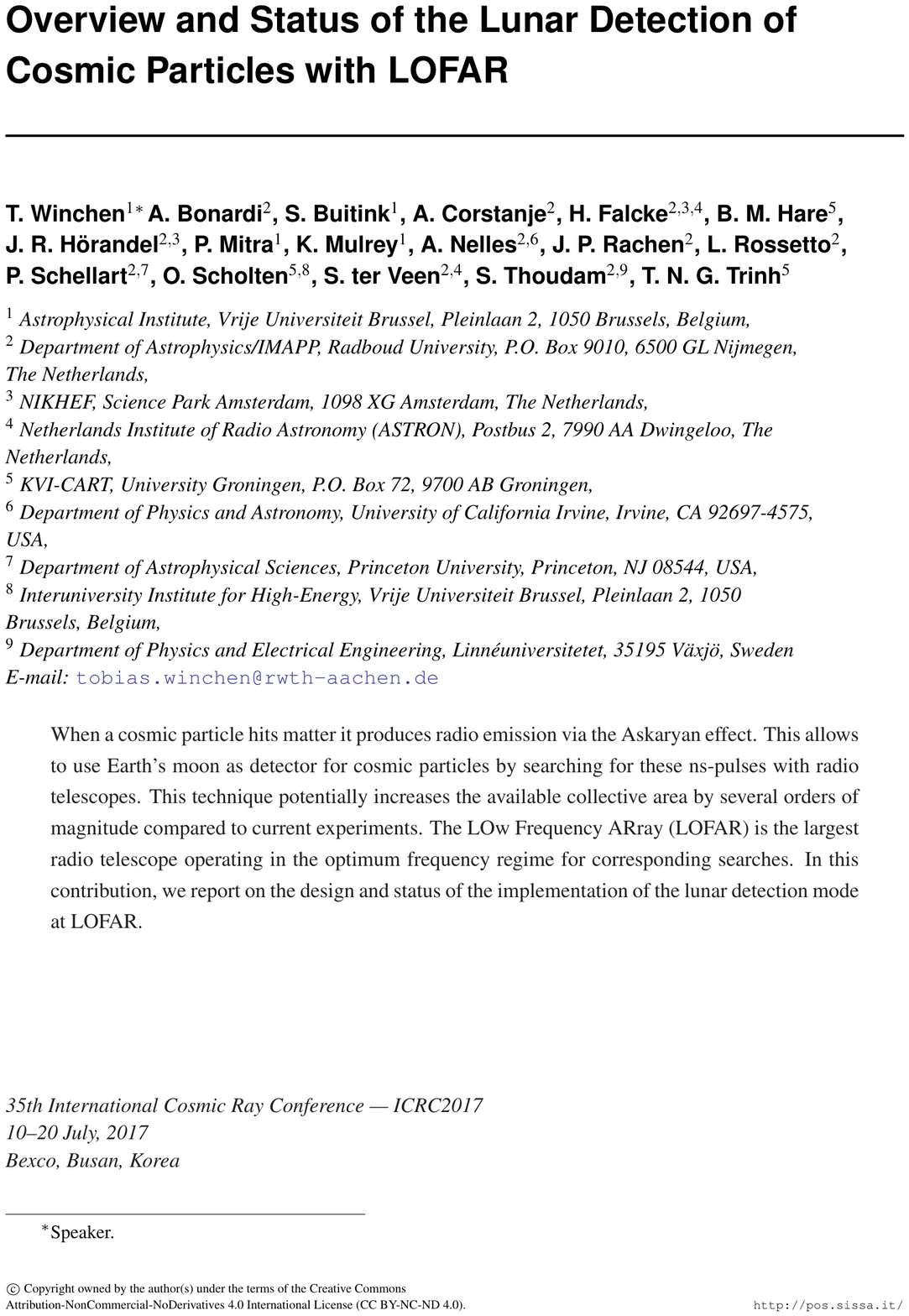}
\includepdf[pages=2-,pagecommand=\thispagestyle{fancy}]{ICRC2017_1061.pdf}
\end{document}